\documentclass[journal]{IEEEtran}
\ifCLASSINFOpdf
\else
\fi
\usepackage{graphicx}
\usepackage{amsmath}

\begin{document}
%
\title{Control and Readout Software in \\Superconducting Quantum Computing}
%
%
%

\author{Cheng Guo, FuTian Liang, Jin Lin, Yu Xu, LiHua Sun, ShengKai Liao, ChengZhi Peng and WeiYue Liu
\thanks{Manuscript received June 12, 2018.}
\thanks{Cheng Guo, Futian Liang, Jin Lin, Yu Xu, Lihua Sun, ShengKai Liao,
and ChengZhi Peng are with the Hefei National Laboratory for Physical
Sciences at the Microscale and Department of Modern Physics, University
of Science and Technology of China, Hefei 230026, China, and Chinese
Academy of Sciences (CAS) Center for Excellence and Synergetic Innovation
Center in Quantum Information and Quantum Physics, University of Science
and Technology of China, Shanghai 201315, China.}
\thanks{WeiYue Liu is with the College of Information Science and Engineering, Ningbo University, Ningbo, China }
\thanks{First author: Cheng Guo (E-mail: fortune@mail.ustc.edu.cn), Corresponding author:
ShengKai Liao (E-mail: skliao@ustc.edu.cn)}
}

\maketitle

\begin{abstract}
The digital-to-analog converter (DAC) and analog-to-digital  converter (ADC) as an important part of the superconducting quantum computer are used to control and readout the qubit states. The complexity of instrument manipulation increases rapidly as the number of qubits grows. Low-speed data transmission, imperfections of realistic instruments and coherent control of qubits are gradually highlighted which have become the bottlenecks in scaling up the number of qubits. To deal with the challenges, we present a instrument management solution in this study. Based on client-server (C/S) model, we develop two servers called Readout Server and Control Server for managing self-innovation digitizer, arbitrary waveform generator (AWG) and ultra-precision DC source which enable to implement physical experiments rapidly. Both Control Server and Readout Server consist three parts: resource manager, waveform engine and communication interface. The resource manager maps the resources of separate instruments to a unified virtual instrument and automatically aligns the timing of waveform channels. The waveform engine generates and processes the waveform for AWGs or captures and analyzes the data from digitizers. The communication interface is responsible for sending and receiving data in an efficient manner. We design a simple data link protocol for digitizers and a multi-threaded communication mechanism for AWGs. By using different network optimization strategies, both data transmission speed of digitizers and AWGs reaches hundreds of Mbps through a single Gigabit-NIC.
\end{abstract}


%
\IEEEpeerreviewmaketitle

\section{Introduction}
%
%
%
%
\IEEEPARstart{Q}{uantum} computer can solve the problem that classical computers can not accomplish attributable to its parallel acceleration characteristics. For the problem of factoring the product $N$ of two large primes, the algorithm complexity of the classical computer using best currently known algorithm\cite{7} requires sub-exponential time $O(e^{{(\log N)}^{1/3}{(\log \log N)}^{2/3}})$, and the algorithm complexity of quantum computer using Shor's algorithm\cite{8} requires only polynomial time $O({(\log N)}^3)$. Benefit from same manufacturing process used in the traditional semiconductor industry, superconducting quantum chips have good scalability, show promise in achieving a high degree of integration and recieve the attention of some technology companies. Google announces a 72-qubit superconducting quantum computing chip called Bristlecone at the APS March 2018 conference. IBM launchs an online quantum computing platform called IBM Q Experience. Intel shows its new 49-qubit superconducting quantum test chip named Tangle Lake at CES 2018. Meanwhile, high level quantum programming languages have been developed. QCL\cite{1}, Quipper\cite{2}, OpenQASM\cite{3}, the advent of these quantum programming languages sketches out a picture of the quantum computer\cite{4}. However, there is a gap between quantum programming languages and quantum chips. Superconducting quantum computing requires precise timing control over a large number of channels. Traditional commercial AWG cannot meet the requirements of customization and rapid application change. In addition, the high price of traditional commercial AWG limits its ability on large-scale application.

\begin{figure}[ht]
\centering 
\includegraphics[scale=0.7]{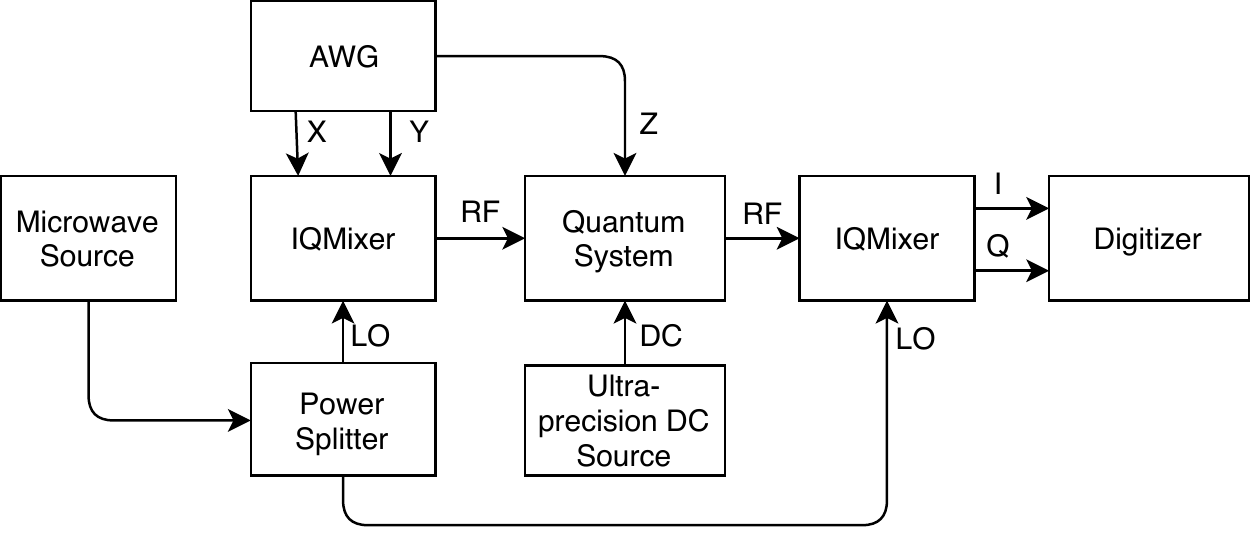}
\caption{Hardware connection for single-qubit controlling}
\label{fig:Hardware}
\end{figure}

In superconducting quantum computing, the state of the qubit is controlled through modulated microwave. In our experiment, the hardware connection is shown in Figure \ref{fig:Hardware}, a qubit needs four independent control channels and two multiplexed readout channels named X, Y, DC, Z, I and Q. The Z and DC are used to provide the bias condition of the qubit. The X and Y are IF signal directly synthesised by AWG, and there exsist ${\pi/2}$ phase difference between X and Y. IQMixer acts as a man-in-the-middle to upconvert controlling IF signals and downconvert RF readout signal. The bias of X and Y leads to the leakage of local frequency. The imbalance of gain and phase leads to the leakage of image frequency. The accuracy of the control signal is related to the microwave detuning which determines the upper limit of fidelity\cite{5}. The imbalance of IQ conversion is an issue that must be considered. Considering bunch of instruments such as digitizer, AWG, ultra-precision DC source, microwave source, etc involved in superconducting quantum computing. As qubit experiments scale from single-qubit to multi-qubit, the complexity of instrument management increases rapidly. Managing these instrument resources flexibly is an important step in building a real quantum computer. To achieve flexible instrument management, it is necessary to develop a software platform. In this paper, we design a instrument management software for Taking a step toward creating a quantum computer.

\section{Layering and modularization}
\begin{figure}[ht]
\centering 
\includegraphics[scale=0.7]{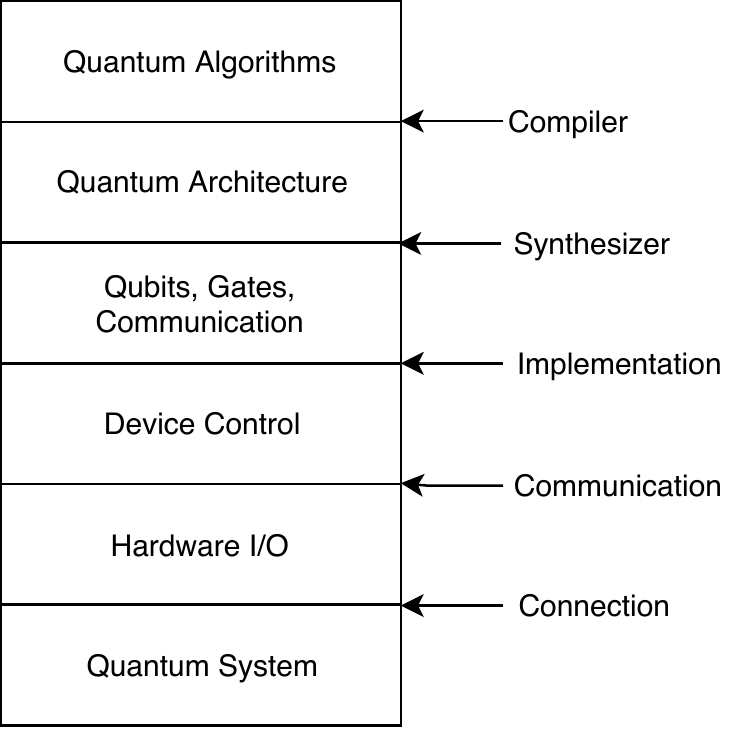}
\caption{Layering of superconducting quantum computing}
\label{fig:Architecture}
\end{figure}

In software engineering, the number of man-months increases dramatically as the number of developers increases\cite{6}. Layering and modularization have proven to be an effective way to overcome software development difficulties in the past. Analogy to classic computers, we summarize the superconducting quantum computer structure as shown in Figure \ref{fig:Architecture}. The compiler is responsible for transforming the high level quantum algorithm to proper code in specified architecture. The function of synthesizer is making architecture specified code to logic qubit operation. Just like field programmable gate array (FPGA), the implementation translates qubit operations to hardware control instructions under the constraint of connections. The communication is used to maximize the data transmission efficiency of device. For hardware connection, it is used to optimize the quality of the signal. 

\begin{figure}[ht]
\centering 
\includegraphics[scale=0.6]{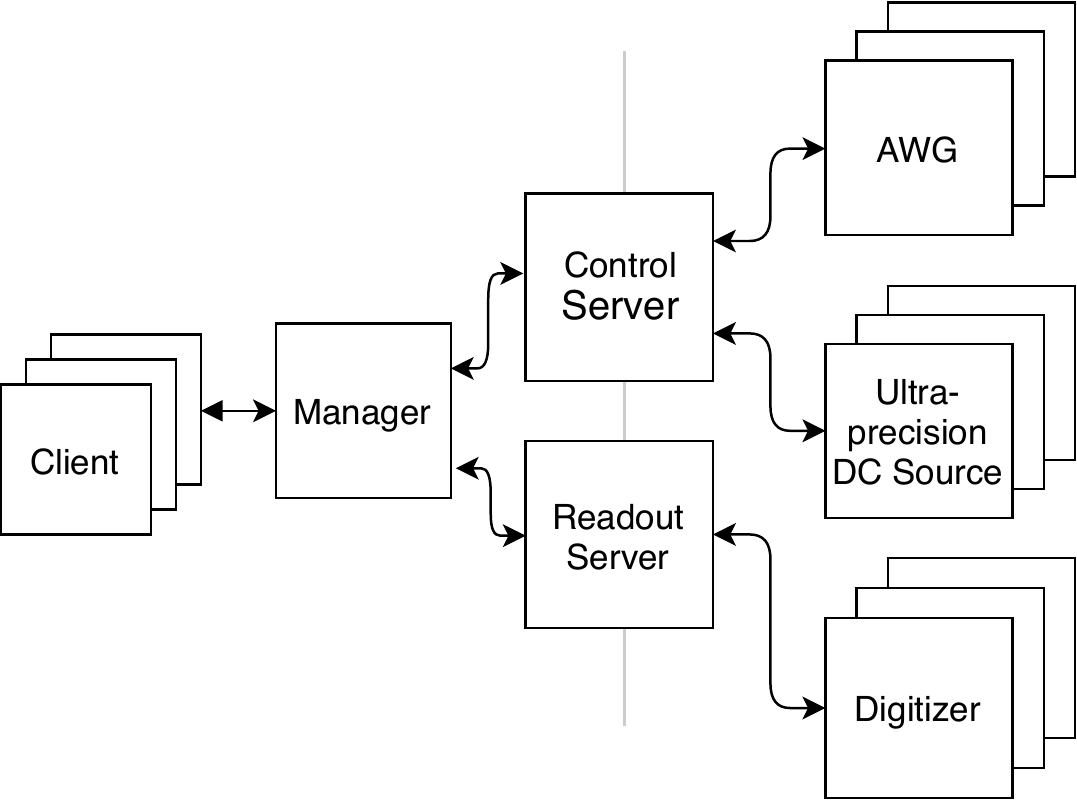}
\caption{Network structure of software}
\label{fig:Network}
\end{figure}

Modularization makes it available for different developers to independently maintain the code which have an advantage when software becomes complicated. As shown in Figure \ref{fig:Network}, we design two stand-alone servers called Control Server and Readout Server based on the idea of modularization. Control Server and Readout Server are in the device control layer. Manager is used to forwarded data between servers and clients. The waveform generation, calculation and analysis are deployed on the server. The client can easily access the service with the remote procedure call (RPC) protocol through manager. The users need only to develop a lightweight client to complete the experiment, which meets the actual needs of rapid development in physical experiments. Taking into account that the data transmission between the clients and servers needs to be forwarded by Manager, transmitting only instructions and calculation results can avoid massive data exchange, reducing the data forwarding pressure of Manager.

\section{Control server}

In order to provide consistent interface, the resource manager of Control Server abstracts a virtual instrument with corresponding resources from AWGs and ultra-precision DC sources regardless of the fact that the devices are physical separate. The virtual instrument have configurable waveform output channels, trigger output channels and DC output channels, each channel can be configured individually.

\begin{figure}[ht]
\centering 
\includegraphics[scale=0.6]{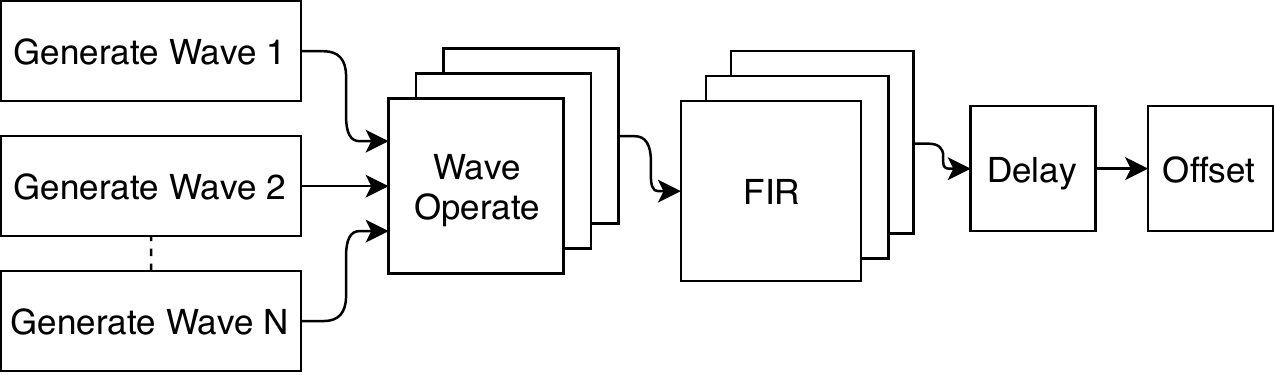}
\caption{Data process diagram of Control Server}
\label{fig:DACData}
\end{figure}

\subsection{Waveform process}

Control Server can store a fixed number of waveforms, and we use the index number to address the waveforms. Like the ALU in CPU, the waveform engine can generate sine, gauss, square, etc and provide addition, multiplication, integral, differential and so on. As shown in Figure \ref{fig:DACData}, after the waveform operations, there are several steps required before outputting the wave, which are designed to overcome the defects of hardware. Z square pulse signal is used to control flux bias of superconducting quantum interference device (SQUID). The raising and falling edges are important in this scenario and we can use finite impulse response (FIR) filter to optimize the edges. For X and Y signals, they are connected to the IQ mixer to acquire RF signal, and the signals need to be calibrated to suppress local frequency and image frequency. For this application, we can configure delay and offset of the channel to solve the problem. By Individually configuring FIR filter, offset and delay of each channel, we can use waveform channels for different purposes. 

The waveform engine supports both symbolic and numerical computation and symbolic computation is implemented based on the GiNaC library. Some calculations, such as convolutions and integrals, are difficult for symbolic computation. To solve the problem, we provide some build-in functions. By using the build-in functions, we do not need to calculate waveform every time to generate the desired waveform. The flattop is a build-in waveform, It is equivalent to the convolution of a Gaussian function with a rectangular pulse.
\begin{table}  
\caption{Time spent on generating waveform}
\label{tabel:time}
\begin{center}
\begin{tabular*}{5cm}{ll}  
\hline  
Waveform  & Time Budgets (us) \\  
\hline
DC & 30\\
Sine & 48 \\  
Rectangle  & 33 \\
Gaussian & 66\\
Isosceles Trapezoid & 46\\
Triangle &47\\
Slope &32\\
Flattop &79\\
\hline  
\end{tabular*}
\end{center}  
\end{table}

In the experiment of randomized benchmarking of quantum gates\cite{10}, waveforms need to be generated frequently. The time spent on waveform generation directly affects the efficiency of the experiment. We test the time it takes to generate waveforms with a waveform length of 6000 points. The test is on a computer with 16GB memory, intel i7-3770k chipset and running Windows7 operating system. The results is shown in Table \ref{tabel:time}.

\subsection{Communication}
\begin{figure}[ht]
\centering 
\includegraphics[scale=0.7]{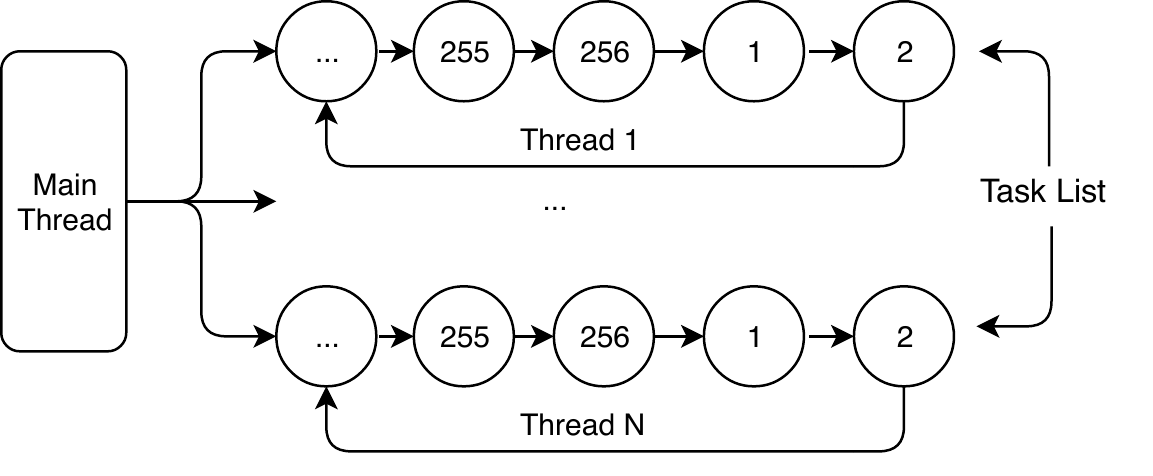}
\caption{Multi-threaded work diagram}
\label{fig:DACThread}
\end{figure}

Both AWGs and ultra-precision DC sources use reliable TCP/IP protocol to communicate. The maximum data transmission speed of AWGs is only several Mbps, and serial communication is unacceptable when there is a lot of data to communicate. In order to make good use of bandwidth, Control Server adopts multi-threaded asynchronous communication scheme. The multi-threaded scheme is shown in Figure~\ref{fig:DACThread}, Control Server maintains an individual task thread for each AWG. The thread cyclically executes the tasks fed by main thread and the execution results are stored in corresponding position for later reading.

\begin{figure}[ht]
\centering 
\includegraphics[scale=0.6]{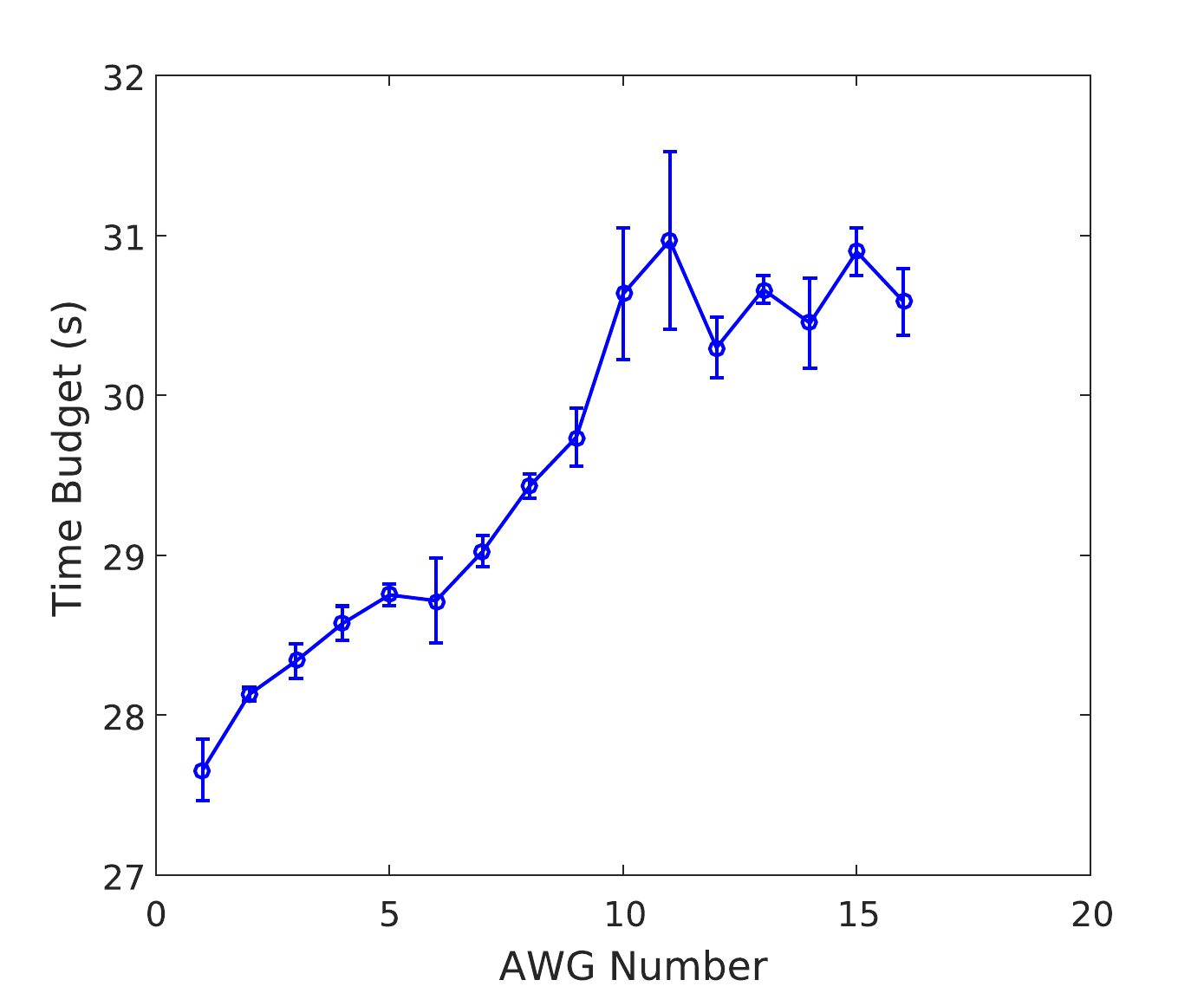}
\caption{Transmission time versus AWG number}
\label{fig:DAC_speed}
\end{figure}

To test the transmission speed of AWGs, we send 25.6 megabyte data to each AWG and measure the time it takes to complete the task. The test result is shown in Figure \ref{fig:DAC_speed}, with the number of AWGs increases, only little extra time needed to finish the task. The multi-threaded scheme improves overall transmission efficiency of the AWGs.

\section{Readout Server}
Like Control Server, Readout Server is used to provide consistent interface of digitizers. Control also provides a virtual digitizer instrument. The virtual instrument have signal input channels and trigger input channels. Each signal input channel can be configured individually.
 
\subsection{Waveform analyse}
\begin{figure}[ht]
\centering 
\includegraphics[scale=0.7]{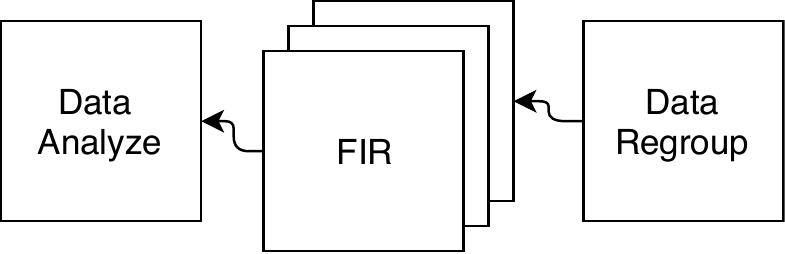}
\caption{Data process diagram of Readout Server}
\label{fig:ADCData}
\end{figure}

As shown in Figure \ref{fig:ADCData}, Readout Server firstly combines the data frames captured from the Ethernet, then pre-processes data using FIR filters ,next analyzes data with a giving algorithm and finally gives a result. 

In dispersive measurement, the phase of readout signal is affected by the qubit state\cite{11}. We use homedyne algorithm to calculate the I, Q value at specified frequency and then expresses the I, Q as a point in IQ plane. By calculating states distribution of $ | 0 \rangle $ and $| 1 \rangle $  in the IQ plane, we can acquire a linear equation. Using this linear equation as a given criterion and we can estimate the state of qubit in next test.

\subsection{Communication}

The digitizers produce data as fast as 1 Gsps per channel, and transmit data through a custom defined data link protocol. We use WinPcap library on PC to capture the packets. By using this method, the data transmission speed can reach more than 400 Mbps through a single Gigabit-NIC. In our experiment, we use interval trigger signal to start the digitizer to acquire data, the length of the data is usually within 10 thousand points, so we can achieve real-time transmission when trigger interval reaches hundreds of microseconds.

\section{Conclusion}
Superconducting quantum computing is a promising scheme to realize quantum computer which receives wide attention. However, there are many difficulties in using traditional commercial instruments to control qubits. Scaling up the number of qubits calls for new hardware and software. This paper introduces a software for instrument control. The software now consists Readout Server and Control Server which are used to provide unify interface of instruments. The waveform engine of Control Server generates common waveforms with length of 6000 points within 100 us, which satisfies the efficiency requirement of randomized benchmarking experiment. The data transmission speed of Readout Server and Control Server reaches hundreds of Mbps, which guarantees the overall efficiency of hardware. In the future, the software needs to be further expanded in terms of  breadth and depth. For breadth, the software can extend new features and new servers. As for depth, a step changes from device control to gate operation is inevitable.

\ifCLASSOPTIONcaptionsoff
  \newpage
\fi



%

\newpage

%

%






\end{document}